\documentstyle[prd,aps,axodraw,
preprint,tighten
]{revtex}

\begin{document}

\newcommand{\TeV}{\,{\rm TeV}}
\newcommand{\GeV}{\,{\rm GeV}}
\newcommand{\MeV}{\,{\rm MeV}}
\newcommand{\keV}{\,{\rm keV}}
\newcommand{\eV}{\,{\rm eV}}
\def\ap{\approx}
\newcommand{\bea}{\begin{eqnarray}}
\newcommand{\eea}{\end{eqnarray}}
\def\beq{\begin{equation}}
\def\eeq{\end{equation}}
\def\haf{\frac{1}{2}}
\def\lpp{\lambda''}
\def\ccg{\cal G}
\def\slash#1{#1\!\!\!\!\!\!/}
\def\u{{\cal U}}

\setcounter{page}{1}
\draft
\preprint{SCIPP-00/08, KAIST-TH 00/06, KIAS-P00015, hep-ph/0004101}

\title{Fermion Electric Dipole Moments in Supersymmetric Models
with R-parity Violation}
 
\author{Kiwoon Choi$^{a,b}$, Eung Jin Chun$^c$ and
Kyuwan Hwang$^b$}

\address{$^a$Santa Cruz Institute for Particle Physics, Santa Cruz,
CA 95064, U. S. A. \\
$^b$Korea Advanced Institute of Science and Technology,
        Taejeon 305-701, Korea \\
$^c$Korea Institute for Advanced Study, 
          Seoul 130-012, Korea \\}


\maketitle

\begin{abstract}
We analyze the electron and neutron electric dipole moments
induced by R-parity violating interactions in supersymmetric models.
It is pointed out that  dominant contributions  can come from
one-loop diagrams involving both the bilinear and trilinear
R-parity odd couplings, leading to somewhat severe constraints
on the products of those couplings.

\end{abstract}



\section{Introduction}

Fermion  dipole moments provide severe constraints on
possible new physics beyond the Standard Model.   
The foremost example is the electric dipole moment (EDM) 
of the electron or neutron whose current experimental bound is
given, respectively, by \cite{eedm,nedm}
\begin{eqnarray} \label{edmexp}
&& |d_e|\leq 4.3\times10^{-27}\, e{\rm cm},  \nonumber\\
&& |d_n|\leq 6.3\times10^{-26}\, e{\rm cm}.
\end{eqnarray}
In the Minimal Supersymmetric Standard Model (MSSM),
apart from the CKM phase,  CP violation can arise from additional 
complex parameters describing the soft supersymmetry (SUSY) breaking
terms in the effective Lagrangian.
Such complex soft parameters generate a light fermion EDM
through one-loop diagrams involving  sparticles (squarks/sleptons
and/or gauginos).  The current bounds on the neutron and electron
EDM constrain the phases of soft parameters to be less than 
${\cal O}(10^{-2})$, or the sfermion masses to be larger than 
${\cal O}(1)$ TeV \cite{mssm} unless there are cancellations 
among various contributions \cite{cancellation}.

When R-parity conservation is not assumed, the MSSM allows 
additional lepton number (L) or baryon number (B)
violating interactions.
In this case, one usually invokes other type of 
(either unbroken or spontaneously broken) symmetry which would ensure 
that  R-parity violating couplings are small enough to avoid too fast
proton decay and too large neutrino masses \cite{symmetry}.
Imposing a symmetry which would forbid (or highly suppress)
B-violating  couplings while allowing non-zero L-violating
couplings is an attractive possibility
since then an interesting pattern of neutrino masses and mixing 
can be generated \cite{rpnumass}.
Furthermore these couplings  
can provide a new source of CP violation, and thus the EDMs of light
fermions \cite{tata,dreiner}.

The quark and lepton EDMs arising from
the R-parity violating Yukawa couplings 
have been studied recently  
in Refs.~\cite{tata,dreiner}.  To construct  effective dipole moment
operator, it is required to have {\it not only} the chirality-flip 
{\it but also} the insertion of even number of R-parity violating couplings. 
In other words, the resulting fermion EDMs 
involve some powers of $\lambda\lambda$ and/or $\lambda\lambda^*$
as well as some  fermion masses or chirality-flipping
Yukawa couplings \cite{barr} where $\lambda$ denotes a generic R-parity
violating Yukawa coupling. At one-loop level,  EDM diagrams involving
$\lambda\lambda$ exist. However such diagrams are suppressed by additional
Majorana  neutrino (or sneutrino) mass 
which would  compensates the $L=2$ of $\lambda\lambda$ in the
EDM coefficients, and thus are negligibly small.
It has been verified that important contributions
can  arise from two-loop diagrams  involving
the insertion of  quark or lepton mass inside the loop 
and also the products of 
R-parity violating couplings such as $\lambda\lambda^*$ \cite{tata} or 
$\lambda^2\lambda^{*2}$ \cite{dreiner}.

\medskip

In this paper, we wish to point out that sizable EDMs
can arise at one-loop level in the presence of both
the bilinear and trilinear R-parity violating interactions.
Let us recall that  generic R-parity violating superpotential
includes the bilinear Higgsino-lepton mixing term
\beq \label{WRp}
W \ni \mu_i L_iH_2,
\eeq
where $L_i$ ($i=1,2,3$) are the lepton $SU(2)$ doublets,
and $H_2$ is the Higgs doublet
with hypercharge $Y=1/2$.
There exist also R-parity violating scalar bilinears in the 
soft SUSY breaking potential:
\begin{equation} \label{VRp}
 V \ni  B_i L_i H_2 + m^2_{L_iH_1} L_i H_1^\dagger,
\end{equation}
where $H_1$ is the Higgs doublet with $Y=-1/2$.
These bilinear scalar terms induce nonzero vacuum expectation values (VEVs)
of sneutrino fields:
\beq \label{svev}
{\langle \tilde{\nu}_i \rangle}
  \propto  B_i^*\langle H^0_2\rangle+ m^{2*}_{L_i H_1}\langle H^0_1\rangle \,.
\eeq
The presence of these bilinear couplings
yields a  mass mixings between
the R-parity even and odd particles,
which are described by the sneutrino VEVs $\langle \tilde{\nu}_i
\rangle$ and the Higgsino-lepton
mixing parameter $\mu_i$.
Since  they  are R-parity odd and also flips
the fermion  chirality without involving additional particles
in the interaction vertex, these mass mixings are more efficient than
the R-parity violating Yukawa coupling $\lambda$ for generating 
fermion EDMs.
We will see that there exist one-loop diagrams for fermion EDMs
which are proportional to the products of bilinear and trilinear
R-parity odd couplings, e.g.  $\langle\tilde{\nu}_i\rangle\lambda$ or 
$\mu_i^* \lambda$.
Such diagrams are {\it not} suppressed by  additional powers
of small fermion masses or Yukawa couplings,
and thus can provide  strong limits on these product of
R-parity odd couplings.

The organization of this paper is as follows.
In Section II, we define the basis on which our analysis is performed
and summarize how to diagonalize the mixings between the R-parity even
and odd particles which are induced by the bilinear R-parity violation.
In Section III, we classify and compute the main contributions 
to the quark and lepton EDMs involving both the bilinear and trilinear
R-parity violating couplings.
The calculational details are presented 
in the Appendix, and the  bounds on couplings 
from the electron and neutron EDMs  are
summarized in Tables 1 and 2 under the assumption that
those couplings take generic complex values.
We conclude in Section IV.

\section{particle mixings induced by  $R$ parity violation}

For comparison with experiments, it is convenient
to work in the quark and lepton mass eigenbasis.  
In this prescription, we
leave the neutrinos in the charged lepton mass eigenbasis since
the mechanism of neutrino mass generation is quite different from 
that of other fermions.  R-parity violation under consideration can be
a major source of nonzero neutrino masses, and then 
the charged lepton mass eigenbasis is particularly
useful for analyzing neutrino mass matrix.  
In the quark and charged lepton mass eigenbasis,
the full superpotential of the MSSM fields including generic
R-parity violating  couplings is given by
\bea 
W &=& \mu H_1 H_2 + h^e_i H_1 L_i E_i^c + 
      h^d_i (H_1^0 D_i D_i^c - H_1^- V^\dagger_{ij} U_j D^c_i)
      + h^u_i(H_2^0 U_i U_i^c -H_2^+ V_{ij} D_j U_i^c) \nonumber \\
&+&  \epsilon_i \mu L_i H_2 + \haf{\lambda}_{ijk}L_i L_j E_k^c 
+ {\lambda}'_{ijk} (L_i^0 D_j D_k^c -  E_i V^\dagger_{jl} U_l D^c_k) 
+ \haf\lambda''_{ijk} U^c_i D^c_j D^c_k \,,
\label{WZ}
\eea
where $L_i$ and $Q_i$ are the lepton and quark $SU(2)$ doublets,
and $E_i^c$, $U_i^c$, $D_i^c$ are the $SU(2)$ singlet
lepton and anti-quark superfields.
Here $V_{ij}$ is the CKM matrix of quark fields, and 
the R-parity violating lepton-Higgsino mass mixing is described by
dimensionless parameter $\epsilon_i$ in the unit of
the conventional Higgsino mass $\mu$.
{}From the above superpotential, one obtains the following 
lepton number violating Yukawa vertices in the Lagrangian;
\begin{eqnarray} \label{LL}
-{\cal L}_{L\!\!\!/} &=& \lambda_{ijk} \left( 
   \tilde{\nu}_{iL} \bar{e}_{kR}e_{jL}
 + \tilde{e}_{jL} \bar{e}_{kR}\nu_{iL}
 + \tilde{e}_{kR}^* \bar{\nu}_{iR}^C e_{jL} \right) \nonumber \\
&+& \lambda'_{ijk}\left( \tilde{\nu}_i \bar{d}_{kR}d_{jL} 
 + \tilde{d}_{jL} \bar{d}_{kR}\nu_{iL}
 + \tilde{d}_{kR}^* \bar{\nu}_{iR}^C d_{jL} \right) \nonumber \\
&-& \lambda'_{ilk}V^\dagger_{lj} \left(
   \tilde{e}_{iL} \bar{d}_{kR}u_{jL}
 + \tilde{u}_{jL} \bar{d}_{kR}e_{iL}
 + \tilde{d}_{kR}^* \bar{e}_{iR}^C u_{jL} \right)\,, \nonumber 
\end{eqnarray}
where the superscript $C$ denotes the charge conjugation.
In order to ensure the longevity of  proton, the  products  
$\lambda' \lambda''$ have to be highly suppressed  \cite{ptri}.
Bilinear R-parity violating terms can also contribute  to proton decay
through the products 
$\langle\tilde{\nu}_i\rangle\lambda''$ or $(\epsilon_i \mu)^* \lambda''$
\cite{pbi}.
In this paper, we simply assume that $\lambda''$ is small enough to
ensure the proton stability, while the L-violating
couplings can take arbitrary values as long as they satisfy
other phenomenological constraints.

\medskip

So far we have defined the quark and lepton mass eigenbasis
without including the effects of R-parity violation.
In the presence of bilinear R-parity violation, there appear
mixings between the ordinary fermions (Higgses) and the
gauginos/Higgsinos (sleptons). As a result, further diagonalization
is required  to define the true mass eigenstates.
In the below, we present the mixing matrices between the R-parity even and
odd particles relevant for our calculation of EDMs.
For this, we generalize the results in Ref.~\cite{kang} to take into
account the complexity of the bilinear parameters in 
Eqs.~(\ref{WRp},\ref{VRp},\ref{svev}), which leads to a mixing 
between the CP even and CP odd neutral Higgses and sneutrinos.  
The process of diagonalization of
slepton-Higgs mixing involves a rotation by which the Goldstone modes
in the Higgs-slepton sector are decoupled \cite{kang}.  
For the sake of simplicity,
we go to the new basis in which the lepton and $Y=-1/2$
Higgs superfields are redefined so that  sneutrinos have vanishing 
VEVs \cite{yuval}.
When the sneutrino VEVs in the original basis are small enough,
i.e.  $a_i \equiv \langle \tilde{\nu}_i^* \rangle/\langle H^0_1 \rangle \ll1$ 
in Eq. (\ref{svev}),  the Lagrangian in the new basis can 
be obtained by the field redefinition
\begin{equation}
 L_i = \hat{L}_i +a_i^* \hat{H}_1\,,\quad 
 H_1 = \hat{H}_i -a_i \hat{L}_1 \,.
\end{equation}
Then, the R-parity odd parameters in the new basis  are given by 
\begin{eqnarray} \label{newbasis}
&&\hat{B}_i = B_i - a_i B\,,  \qquad 
\hat{m}^2_{L_iH_1}= m^2_{L_iH_1} + a_i (m^2_{L_i}-m^2_{H_1}) \nonumber \\
&& \hat{\epsilon}_i= \epsilon_i -a_i\,,\quad
\hat{\lambda}_{ijk} = \lambda_{ijk} - a_i h^e_j \delta_{jk}\,, \quad 
\hat{\lambda}'_{ijk} = \lambda'_{ijk} - a_i h^d_j \delta_{jk} 
\end{eqnarray}
where $h^e_j$ and $h^d_j$ are the Yukawa couplings of the charged leptons
and down-type quarks, respectively.  Note that we have the relation 
$\hat{B}_it_\beta + \hat{m}^2_{L_iH_1} = 0$
where $t_\beta \equiv \tan\beta= \langle H_2^0\rangle / \langle H_1^0\rangle$,
which assures that sneutrino VEVs vanish in the new basis.
In the following, we will drop hats to denote quantities 
in this new basis of vanishing sneutrino VEVs.

{\bf  Neutrino-neutralino mixing}:  Let us first consider the well-known
mixing among neutral fermions \cite{rpnumass} for the completeness.
In the basis of vanishing sneutrino VEVs, 
the mass mixing  between the neutrinos ($\nu_i$) and 
the neutralinos $N_I=(\tilde{B}, \tilde{W}_3, \tilde{H}_1^0, \tilde{H}^0_2)$
are described by the following $3\times 4$ matrix:
\begin{equation} \label{MnN}
\left({\bf M_D}\right)_{iI} = \left( 0, 0, 0, - \epsilon_i \mu \right)\,.
\end{equation}
Combined with the usual $4\times 4$ neutralino mass matrix ${\bf M_N}$
of $N_I=(\tilde{B}, \tilde{W}_3, \tilde{H}_1^0, \tilde{H}_2^0)$,
the above $3\times 4$ mixing matrix ${\bf M_D}$ forms a $7\times 7$
mass matrix of the neutrinos and neutralinos.
At leading order in the expansion in powers of small ${\bf M_D}$,
the $7\times 7$ neutral fermion mass matrix can be
diagonalized by the rotation:
\begin{equation} \label{UN}
 {\cal U}^N \equiv 
  \pmatrix{ 1 & \Theta^N \cr -(\Theta^{N})^T & 1}
  \pmatrix{ U^\nu & 0 \cr 0 &  N}
\end{equation}
where $\Theta^N \equiv {\bf M}_{D}{\bf M}^{-1}_N$, $U^\nu$ is the $3\times
3$ neutrino mixing matrix, and finally $N$ is the usual $4\times 4$ 
diagonalization matrix of ${\bf M_N}$ in the absence of R-parity violation 
\cite{haber}.  
Here the mixing matrix $\Theta^N$ is given by \cite{nowa}
\begin{equation} \label{TN}
 \Theta^N_{iI} = \epsilon_i c^N_I c_\beta + \epsilon_i\delta_{I3} 
\eeq
with
\beq
  c^N_I= {M_Z\over F_N}
   \left(s_W{M_2\over M_{\tilde{\gamma}}}, 
       -c_W{M_1\over M_{\tilde{\gamma}}}, 
        s_\beta{M_Z\over \mu},
      -c_\beta{M_Z\over\mu} \right),
\end{equation}
where  
$F_N\equiv  M_1M_2/M_{\tilde{\gamma}} - M^2_Z s_{2\beta}/\mu$,
$M_{\tilde{\gamma}}\equiv c_W^2 M_1 + s_W^2 M_2$
for the $SU(2)\times U(1)$ gaugino mass parameters $M_1$, $M_2$,
$s_W=\sin\theta_W$ with $\theta_W$ being the weak mixing angle,
and $s_\beta=\sin\beta$, {\it etc}.
In the above, for the sake of simplicity,  we assume that 
$M_1$, $M_2$, and $\mu$ are approximately real, which may be necessary
also to avoid  a too large neutron EDM.

{\bf Charged lepton-chargino mixing}: 
R-parity violation leads also a mixing between  the charged
leptons and the charged gaugino/Higgsino.  Following the steps 
similar to the case of neutral fermions,
one finds that the $5\times 5$ diagonalization matrices 
${\cal U}^L,{\cal U}^R$ which would bring 
$(e_i, \tilde{W}^-, \tilde{H_1}^-)$  and 
$(e^c_i, \tilde{W}^+, \tilde{H_2}^+)$ to the mass eigenstates
$(e^{\pm}_i,\chi^{\pm}_1,\chi^{\pm}_2)$  
are given by
\begin{equation} \label{ULR}
{\cal U}^{L,R} =
 \pmatrix{ 1 & \Theta^{L,R} \cr -\Theta^{L,R\dagger} & 1\cr}
 \pmatrix{1 & 0 \cr 0 & L,R\cr} 
\end{equation}
where 
\beq \label{TLR}
 \Theta^L_{ia} = \epsilon_i c^L_a c_\beta + \epsilon_i \delta_{a2}\,,
 \quad  \Theta^R_{ia} = {m^e_i \over F_C}\epsilon_i c^R_a  c_\beta \,,
\eeq
with 
\begin{eqnarray} 
 c^L_a &=&  -\left( \sqrt{2}{M_W\over F_C},
       2s_{\beta} {M_W^2\over \mu F_C} \right)\,,
\nonumber \\
 c^R_a &=&  -\left( \sqrt{2} {M_W (\mu-M_2 t_\beta) \over \mu F_C}, 
    {(M_2^2+2M_W^2 c^2_\beta)\over c_\beta\mu F_C} \right)
\,. \nonumber
\end{eqnarray}
Here  $F_C\equiv M_2 +M_W^2s_{\beta}/\mu$, and
$m^e_i$ denote the charged lepton mass. 
Note that $\Theta^R_{ia}$ is suppressed by the factor  $m^e_i/F_C$ 
compared to $\Theta^L_{ia}$.
In Eq.~(\ref{ULR}), the $2\times 2$ matrices $L,R$ 
are the usual chargino mass diagonalization
matrices in the absence of R-parity violation \cite{haber},
and the index $a=1,2$ labels the  chargino fields 
before the diagonalization by $L,R$.

{\bf Charged slepton-charged Higgs mixing}:
In the basis with vanishing sneutrinos VEVs,  no mass mixing between
the slepton and Higgses arises from the D-term potential.
Then  the mixing comes from
the following part of the scalar potential:
\begin{equation} \label{SLH}
 V \ni B_i (L_i H_2 -  t_\beta L_i H_1^\dagger )
  + \epsilon_i h^e_i \mu H_2 H_1^\dagger E^{c*}_i + h.c.,
\end{equation}
where we have used  the relation
$m^2_{L_iH_1}=-B_it_\beta$ for vanishing sneutrino VEVs.
Decoupling the Goldstone mode by setting  $H_1^-=s_\beta H^-$ and
$H_2^-=c_\beta H^-$, we find that the mixing mass terms for the charged
slepton and Higgs fields are given by
\begin{equation}
 V \ni -(B_i/c_\beta) \tilde{e}_{iL} H^+ 
       -(m^e_i\mu/c_\beta) \tilde{e}_{iR} H^+ + h.c.,
\end{equation}
where $\tilde{e}_{iL}$ and $\tilde{e}_{iR}$ are the left-handed  and 
right-handed  sleptons, respectively.
Applying again the seesaw diagonalization  as in the case of 
Eqs.~(\ref{UN},\ref{ULR}), one finds
the $8\times 7$ rotation matrix ${\cal U}^C$ 
which brings the charged scalar fields
($\tilde{e}_{iL}, \tilde{e}_{iR}, H_1^-, H_2^{-}$) to the mass eigenstates
($\tilde{e}_{i1}, \tilde{e}_{i2}, H^-$):
\begin{equation} \label{UC}
 {\cal U}^C =
 \pmatrix{ 1 & 0 &  \eta^i_1 \cr
           0 & 1 &  \eta^i_2 \cr
          -\eta^{i*}_1 s_\beta & -\eta^{i*}_2 s_\beta  & s_\beta \cr 
      -\eta^{i*}_1 c_\beta & -\eta^{i*}_2 c_\beta & c_\beta \cr} \,,
\end{equation}
where  
\begin{eqnarray} \label{etas}
 \eta^i_1 &=& {B_i(M^2_{Ri}-m^2_{H^-}) - \epsilon_i m^e_i \mu M^2_{Di} \over
    c_\beta (m^2_{\tilde{e}_{i1}}-m^2_{H^-}) 
            (m^2_{\tilde{e}_{i2}}-m^2_{H^-}) } \,, \nonumber \\
 \eta^i_2 &=& { \epsilon_i m^e_i \mu (M^2_{Li}-m^2_{H^-}) - B_iM^2_{Di} \over
    c_\beta (m^2_{\tilde{e}_{i1}}-m^2_{H^-}) 
            (m^2_{\tilde{e}_{i2}}-m^2_{H^-}) }  \,,  \nonumber
\end{eqnarray}
for $M^2_{Li}$, $M^2_{Ri}$ and $M^2_{Di}$ being the  
left-left, right-right and left-right mass terms of slepton fields
in the case of R-parity conservation, respectively.
Here we have neglected the diagonalization of the
slepton mass matrix which is assumed to be an approximate unit matrix 
for reasonable range of $\tan\beta$.

{\bf Sneutrino-neutral Higgs mixing}:
If we allow $m^2_{L_i H_1}$ and/or $B_i$ to be complex-valued, 
these $R$ parity violating parameters lead to a mixing not
only between the R-parity even and odd states,
but also between the CP even and odd states. 
Then the usual decomposition of  slepton/Higgs sector into
CP-even and CP-odd parts \cite{yuval} cannot be applied any more.
Again decoupling the Goldstone mode by setting
${\rm Im}(H_1^0) = s_\beta A^0$ and 
${\rm Im}(H_2^0) = c_\beta A^0$, we find the mixing mass terms
for $({\rm Re}(\tilde{\nu}_i), 
{\rm Im}(\tilde{\nu}_i))$ and (${\rm Re}(H_1), 
{\rm Re}(H_2), A^0$) are given by
\begin{equation}
\pmatrix{ -{\rm Re}(B_i)t_\beta & {\rm Re}(B_i) & -{\rm Im}(B_i)/c_\beta\cr
          {\rm Im}(B_i)t_\beta & -{\rm Im}(B_i) & -{\rm Re}(B_i)/c_\beta\cr}\,.
\end{equation}
Combining the above mixing mass matrix with the R-parity conserving 
slepton and 
Higgs mass matrix, we get the diagonalization matrix given  as follows:
\begin{equation} \label{US}
 {\cal U}^S =
\pmatrix{  1 & \Theta^{S}_i \cr -(\Theta^{S}_i)^T & 1 \cr}
\pmatrix{  1 & 0  \cr 0 & U_\alpha  \cr }
\end{equation}
where the index $i$ labels each slepton generation and $U_\alpha$ is 
a block-diagonal $3\times 3$ matrix in which the upper-left block 
is a $2\times2$ matrix diagonalizing the CP-even Higgs bosons 
in the absence of R-parity violation \cite{haber} and the 
lower-right block is identity.
Here we keep using the standard notation for the Higgs-like scalar fields
although they are not exact CP eigenstates.
Again, the slepton diagonalization part is neglected.
In Eq.~(\ref{US}), $\Theta^S$ (for each slepton flavor $i$)
is a $2\times 3$ matrix whose components are given by
\begin{eqnarray} \label{TS}
\Theta^S_{11} = {(m_{\tilde{\nu}_i}^2 + M_Z^2 c_{2\beta} ) t_\beta 
  {\rm Re}({B}_i) \over F_S}\, ,  \quad
&\Theta^S_{21}& = -{(m_{\tilde{\nu}_i}^2 + M_Z^2 c_{2\beta}) t_\beta 
  {\rm Im} ({B}_i) \over F_S} \nonumber\\
\Theta^S_{12} = -{(m_{\tilde{\nu}_i}^2 - M_Z^2 c_{2\beta} )  
  {\rm Re} ({B}_i) \over F_S}\,,  \quad
&\Theta^S_{22} &= {(m_{\tilde{\nu}_i}^2 - M_Z^2 c_{2\beta}) 
  {\rm Im} ({B}_i) \over F_S} \\
\Theta^S_{13} = { {\rm Im}(B_i) \over 
  c_\beta (m_{\tilde{\nu}_i}^2-m_A^2)  }\,,  \quad
&\Theta^S_{23}& = { {\rm Re}(B_i)  \over
  c_\beta (m_{\tilde{\nu}_i}^2-m_A^2) } \nonumber
\end{eqnarray}
where $F_S \equiv  m_{\tilde{\nu}_i}^2 ( m_{\tilde{\nu}_i}^2
- m_{h^0}^2 -m_{H^0}^2  ) +m_{h^0}^2 m_{H^0}^2$ and 
$m_{\tilde{\nu}_i}^2 = m^2_{L_i} + M_Z^2c_{2\beta}/2$.
As we will see in the Appendix, the EDMs arising from slepton-Higgs mixing
turns out to be proportional to the charged lepton or down-type Yukawa
couplings [see the terms containing ${\cal U}^S$ in Eqs.~(\ref{Ae},\ref{Ad})]
and thus are subleading.

An important consequence of neutrino-neutralino mixing is the
generation of nonzero neutrino mass \cite{rpnumass}.
Diagonalizing   the $7\times 7$ neutral fermion mass matrix, one finds
the  neutrino mass matrix at tree-level:
\beq
m^{\rm tree}_{ij} = {M^2_Z \over F_N} c_\beta^2 \epsilon_i \epsilon_j. 
\label{neumas}
\eeq 
The existing atmospheric and solar neutrino data suggest that
the largest neutrino mass eigenvalue is of the order of 0.1 eV
\cite{skatm}.
Also the neutrino masses induced by R-parity violation  are 
generically hierarchical and dominated by the above tree level mass matrix.
We then have
\beq
\epsilon \equiv (\sum_i |\epsilon_i|^2)^{1/2} \simeq 
0.9\times10^{-6}{1\over c_\beta}
\left({F_N \over M_Z}\right)^\haf 
\left({m_{\nu_3}\over 0.1\eV} \right)^\haf\,,
\eeq
where $m_{\nu_3}$ denotes the largest mass eigenvalue
of (\ref{neumas}).
If R-parity violation is the dominant source of neutrino masses,
the large mixing of atmospheric neutrinos \cite{skatm} would imply 
$\epsilon \approx |\epsilon_2| \approx |\epsilon_3|$, and  
one can have $|\epsilon_1| \approx \epsilon$
or $|\epsilon_1| \approx 0.04 \epsilon$ 
depending on the choice of the large or small mixing angle solution 
to the solar neutrino problem \cite{solan}.

\section{fermion EDM from R-parity violation}

The fermion EDM operator is given by
\beq
\frac{i}{2}d_{\psi}\bar{\psi}\sigma^{\mu\nu}\gamma_5\psi F_{\mu\nu}
\eeq
for a Dirac fermion $\psi$ and the electromagnetic field
strength $F_{\mu\nu}$.
Obviously this EDM operator flips the fermion
chirality and also breaks CP invariance.
It has been argued that there is no one-loop fermion EDM from
R-parity violating trilinear Yukawa couplings 
in the limit of massless neutrino
\cite{tata,dreiner}. 
To generate  R-parity even  EDM, one needs to insert
even number of R-parity odd couplings.
Then it is  not possible to have one-loop EDM diagrams involving the two
insertions of R-parity
violating Yukawa couplings  in a manner compatible
with both the requirement of chirality-flip and the
chiral structure of Yukawa vertices. 
However in the presence of bilinear R-parity violation which would yield
a mixing between the R-parity even and odd particles,
a combination of R-parity violating bilinear and trilinear
couplings can induce a fermion EDM at one-loop level.
An important feature of these one-loop diagrams is that
the second trilinear interaction vertex is provided by gauge interactions
(or the large top quark Yukawa coupling), so
there is no  small fermion mass or Yukawa couplings
involved other than the minimal product of R-parity odd bilinear and trilinear
couplings.
FIGs.~1--3 show such diagrams for the EDMs of charged lepton,
down-type quarks, and up-type quarks, respectively.
One can also have one-loop diagrams involving the two 
insertions of R-parity odd bilinear couplings. However   
such diagrams  are suppressed by small neutrino mass,
and turn out to be negligibly small.

\medskip

Let us first consider the leading contributions to the electron EDM 
in more detail.
The diagrams of FIG.~1 without a flavor change of sleptons
inside the loop, i.e. 
$i=k=\alpha=1$,  give rise to the electron
EDM :
\begin{eqnarray} \label{eedm}
d_{e} = {eg \over 16\pi^2}&& \left\{
{\rm Im}\left( \u^R_{W\chi_n} \u^L_{j\chi_n} \lambda_{1 j 1} \right)
     {1 \over m_{\tilde{\nu}_1}}
 G_f(m_{\chi^\pm_n}; m_{\tilde{\nu}_1}) \right.
        \nonumber\\
&&- {1\over\sqrt{2}} {\rm Im}\left( [\u^N_{W\chi_n}+ t_W \u^N_{B\chi_n}] 
 \u^N_{j\chi_n} \lambda_{1j1} \right)
 {1 \over m_{\tilde{e}_1}}
 G_s(m_{\chi^0_n};m_{\tilde{e}_1}) \nonumber\\
&& +\sqrt{2} t_W  \left.
  {\rm Im}\left(\u^N_{B\chi_n} \u^N_{j\chi_n} \lambda_{1 j 1} \right)
 { 1\over m_{\tilde{e}^c_1}}
 G_s(m_{\chi^0_n}; m_{\tilde{e}^c_1}) \right\} \,,
\end{eqnarray} 
where $G_f$, $G_s$ are the  loop functions defined in the Appendix
and $\u^L,\u^R,\u^N$ are the diagonalization matrices discussed
in the previous section.
Here $\chi^\pm_n$, $\chi^0_n$ are the fermions  inside the loops
of FIG.~1 which can be either  R-parity even,
i.e.  charged lepton or neutrino, or R-parity odd,
i.e. chargino or  neutralino. 
It turns out that the terms with
R-parity even $\chi_n$ is negligible, so we will keep
only the terms with R-parity odd $\chi_n$.
In this case, $\u^R_{W\chi_n}$, $\u^N_{W\chi_n}$, $\u^N_{B\chi_n}$ 
correspond to
the R-parity even elements of
the usual chargino or neutralino diagonalization matrix,
while $\u^L_{j\chi_n}$ or $\u^N_{j\chi_n}$ contains
the R-parity odd  element  $\Theta^N_{jI}$ or $\Theta^L_{ja}$    
which connects the neutrino with neutralino or the charged lepton with
chargino. 

Applying the experimental bound on the electron EDM
to  Eq.~(\ref{eedm}), we can obtain the bounds on the products of
the trilinear and bilinear R-parity violating couplings.
For a rough estimate, it is useful to recall that 
the mass mixing between the charged wino $\tilde{W}^+$ 
and the charged lepton $e_j$ 
(denoted by the cross in the fermion line of FIG.~1a)
is given by $M_W \epsilon_j c_\beta$,  implying 
$$\u^R_{W\chi_n} \u^L_{j\chi_n} m_{\chi_n^\pm} \approx M_W \epsilon_j 
c_\beta.$$
Then applying the experimental bound (\ref{edmexp})
to the first term of Eq.~(\ref{eedm}),
we get
\begin{equation} \label{bound1}
 {\rm Im}\left(\lambda_{1j1}\epsilon_j^* c_\beta\right),\;
 {\rm Im}\left(\lambda_{1j1}\epsilon_j^*{M_W\over m_{\chi^\pm}}\right)  
 < 1.4\times10^{-8} {m_{\chi^\pm} m_{\tilde{\nu}} \over (100 {\rm GeV})^2}\,.
\end{equation}
Here the bound on the product
$\lambda_{1j1}\epsilon^*_jM_W/m_{\chi^{\pm}}$
can be understood by that the mixing mass of $\tilde{H}_2^+$ and 
the charged lepton  is given by $\epsilon_j \mu$ and 
the Wino-Higgsino mixing is of the order $M_W/m_{\chi}$ with $M_W<m_{\chi}$.
For numerical estimates, we took a typical value of the loop functions,
$G_f(m_f; m_\alpha) \approx 1/3$ and 
$G_s(m_f; m_\alpha) \approx 1/6$.
Note that the values of these loop functions are rather insenstive to 
$m_f/m_{\alpha}$ as can be seen in FIG.~4.
The other terms in Eq.~(\ref{eedm}) are expected to yield 
a bit weaker bound by the factor of $t_W$.  
If the dominant source of neutrino masses were the bilinear R-parity violation,
we have $\epsilon c_\beta=(F_N m_\nu)^{1/2}/M_Z$ as discussed in
the previous section.
This enables us to translate the bounds of Eq. (\ref{bound1})
into a bound on the R-parity violating Yukawa coupling:
\begin{equation} \label{bound1s}
 | \lambda_{1j1}|< 0.013 {\epsilon \over |\epsilon_j|}
  \left(m_{\chi^\pm} \over 100 \GeV\right)^{1/2}
  \left(m_{\tilde{\nu}} \over 100 \GeV\right)
  \left(m_\nu \over 0.1 \eV\right)^{-1/2},
\end{equation}
where we assumed that the phases are
of order unity and $F_N\approx m_{\chi^\pm}$.

\medskip

If we allow a generation mixing in the squark or slepton sector, 
we have more contributions to the fermion dipole moments.
To deal with this generation mixing, let us use the so-called mass 
insertion method \cite{scalmix}.  In this approach, 
the amount of flavor violation is parameterized by the quantity,
\beq
\delta^{f, AB}_{ij} \equiv { \Delta^{f,AB}_{ij} \over 
     m_{\tilde{f}_i} m_{\tilde{f}_j} } \,,
\eeq
where $\Delta^{f,AB}_{ij}$ is the off-diagonal element and 
$m^2_{\tilde{f}_i}$ is the diagonal element of the sfermion mass-squared
matrix in the quark and lepton mass eigenbasis,
$f=u,d,l$ are the indices for the up-type squarks, down-type squarks, 
sleptons, and $A,B=L,R$ denote the chirality of these sfermions.
Including the flavor mixing of sfermions, we have for instance
the following contribution to the electron EDM:
\beq
d_e = {eg \over 16\pi^2}
{\rm Im}\left( \u^R_{W\chi_n} \u^L_{j\chi_n} 
\delta^{l,LL}_{1\alpha} \lambda_{\alpha j 1} \right)
 H_f\left(m_{\chi^\pm_n}; m_{\tilde{\nu}_1}, m_{\tilde{\nu}_\alpha}
\right)\,, 
\eeq 
which reproduces the first term in Eq.~(\ref{eedm}) when $1=\alpha$.
The definition of the loop function $H_f$ and also the other contributions
can be found in the Appendix.
Again applying the experimental bound 
to these contributions to the electron EDM,
we get
\begin{eqnarray} \label{bound1f}
&& {\rm Im}\left(\delta^{l,LL}_{1\alpha}\lambda_{\alpha j1}
               \epsilon_j^* c_\beta\right),\;
 {\rm Im}\left(\delta^{l,LL}_{1\alpha}\lambda_{\alpha j1}
               \epsilon_j^*{M_W\over m_\chi}\right) 
 < 1.4\times10^{-8} {m_{\chi} m_{\tilde{l}} \over (100 {\rm GeV})^2}
\nonumber \\
&& {\rm Im}\left(\delta^{l,RR}_{1\alpha}\lambda_{1j\alpha}
               \epsilon_j^* c_\beta\right),\;
 {\rm Im}\left(\delta^{l,RR}_{1\alpha}\lambda_{1j\alpha}
               \epsilon_j^*{M_W\over m_\chi}\right) 
 < 6.4\times10^{-8} {m_{\chi} m_{\tilde{l}} \over (100 {\rm GeV})^2}\,,
\end{eqnarray}
where the first bound comes from the diagram of FIG.~1a or 1b exchanging
$\tilde{\nu}$ or $\tilde{e}$ and the second one from FIG.~1c
exchanging $\tilde{e}^c$.
Most stringent bounds on $\delta^{l,AB}_{ij}$ with $AB=LL,RR$
come from the $l_i\to l_j\gamma$ decays.  
One then finds 
$\delta^{l,AB}_{12} \lesssim 10^{-2} (m_{\tilde{l}}/100\GeV)^2$, however
there are essentially no bounds
on $\delta^{l,AB}_{13}$ and  $\delta^{1,AB}_{23}$ 
\cite{scalmix}.  

\medskip

The most important contributions to the down quark EDM come from 
the gauge vertex diagrams [FIG.~2a, 2b],
and also the top  Yukawa vertex diagrams [FIG.~2c],  from which we find
\bea \label{dedm}
d_{d} &=& 
{e g \over 16 \pi^2 } \left\{ 
 -{\rm Im} \left( \u^L_{j\chi^-_n} \u^R_{W\chi^+_n} 
 V_{\alpha 1}V^{\dagger}_{l\alpha}  \lambda'_{jl1} \right) 
 {1\over m_{\tilde{u}_\alpha} } 
 \left[ {2\over 3} G_s(m^2_{\chi_n}; m_{\tilde{u}_\alpha} ) 
 + G_f(m_{\chi_n}; m_{\tilde{u}_\alpha})  \right]\right. \nonumber\\
&&+ {1\over 3\sqrt{2}} {\rm Im}\left( 
   [\u^N_{W\chi_n}- {t_W\over 3} \u^N_{B\chi_n}] 
 \u^N_{j\chi^0_n} \lambda_{j11} \right) 
 {1\over  m_{\tilde{d}_1} } 
 G_s(m_{\chi^0_n}; m_{\tilde{d}_1} ) \nonumber\\
&&+ \left. {\sqrt{2}\over 9} t_W {\rm Im}\left(\u^N_{B\chi_n} \u^N_{j\chi_n}
 \lambda'_{\j 1 1} \right)
 {1\over  m_{\tilde{d}^c_1} } 
 G_s(m_{\chi^0_n}; m_{\tilde{d^c}_1} )  \right\}\nonumber\\
&-& {e h_t \over 16 \pi^2 } 
 {\rm Im}\left(\u^{C*}_{H_2\phi^-_n} \u^C_{j\phi^-_n} 
 V_{31}V^{\dagger}_{l3} \lambda'_{jl1} \right)
 {1\over  m_{\phi^\pm_n} } 
 \left[ {2\over 3} G_f(m_{t} ;m_{\phi^\pm_n}) 
 + G_s(m_t ;m_{\phi^\pm_n})  \right] \,.
\eea
More general expression including the squark mixing effects is given 
in the Appendix.
{}From the valence quark contribution to the neutron EDM
$$d_n = {1\over 3} (4d_d - d_u) r^e\,,$$
taking into account the renormalization group effect from the weak scale
to the hadronic scale with $r^e\simeq 1.5$  \cite{rgedm},
the current experimental bound (\ref{edmexp}) gives
\begin{equation} \label{bound2}
 {\rm Im}\left(V_{\alpha1}V^\dagger_{l\alpha}
         \lambda'_{jl1}\epsilon_j^* c_\beta, \right),\;
 {\rm Im}\left(V_{\alpha1}V^\dagger_{l\alpha}
         \lambda'_{jl1} \epsilon_j^*{M_W\over m_{\chi}}\right)  
 < 1.0\times10^{-7} {m_{\chi} m_{\tilde{q}} \over (100 {\rm GeV})^2}\,.
\end{equation}
Again if the bilinear R-parity violation gives $m_{\nu}\approx 0.1$
eV, the above bound  would give 
$\lambda'_{j11} \lesssim 0.1$.
{}From the top quark exchange diagram [FIG.~2c] giving the 
last contribution of Eq.~(\ref{dedm}), one finds
$$\u^{C*}_{H_2\phi^-_n} \u^C_{j\phi^-_n} \approx \eta_1^{j*} c_\beta $$
as can be read off from Eq.~(\ref{UC}).
The contribution involving $\eta_2^i$ is neglected as it is proportional
to the lepton mass $m^i_e$.
Then one gets
\begin{eqnarray}
 {\rm Im}\left( V^\dagger_{l3} \lambda'_{jl1} 
          {B_j^*\over m^2_{\tilde{l}}} \right) 
   < 2.5\times10^{-6} \; {0.01\over V_{31}}
	 {m_{\tilde{l}} \over 100 \GeV} \,.
\end{eqnarray}
If we allow the squark generation mixing, 
$V_{\alpha1}V^\dagger_{l\alpha} \lambda'_{jl1}$ 
can be replaced by $\lambda'_{i\alpha 1}\delta^{U,LL}_{\alpha 1}$
in Eq.~(\ref{bound2}).
More general expressions
with both the CKM and squark mixings are given in the Appendix.

\medskip

Contrary to the down quark EDM,  the up quark EDM in our scheme turns out
to be very small as it is further suppresed by small
quark masses.
For instance, the diagram in FIG.~3 gives rise to 
\begin{equation} \label{uedm}
 d_u= {e h_u \over 16\pi^2}
 {\rm Im}\left( \u^L_{i \chi_n} \u^R_{H \chi_n} V_{1\alpha} V^\dagger_{\beta 1}
 \delta^{d,LR}_{\alpha\alpha} \lambda'_{i\beta\alpha} \right)
 {1\over m_{\tilde{d}_\alpha} }
  \left[ -{1\over 3} G_s(m_{\chi^-_n}; m_{\tilde{d}_\alpha}) 
 + G_f(m_{\chi^-_n}; m_{\tilde{d}_\alpha}) \right] 
\eeq
which is the limiting case of $\alpha=\beta$ in Eq.~(\ref{Au}).
One can see from Eq.~(\ref{uedm}) that the up quark EDM 
includes additional suppression factor
$m_u m_d/ m^2_{\tilde{q}}$.

\medskip

The nucleon EDM gets  contributions not only from the
quark EDMs, but also from the chromoelectric dipole moments (CDMs)
of light quarks \cite{cdm}.
Since the diagrams for the CDMs of the up and down quarks
have the same structure as those for the EDMs under the replacement of
the external photon by gluons, one does not find a new bound on the
R-parity violating couplings.  On the other hand, 
there is also a contribution
from the strange quark CDM whose diagrams involve a different set of
R-parity violating couplings.  The strange quark CDM ($d_s^c$)  
contribution to
the neutron EDM is given by $d_n \approx 0.027 d^c_s r^c$ \cite{pak}
with $r^c\simeq 3.3$ coming from  the renormalization group effect, 
which leads to 
\begin{equation} \label{bound3}
 {\rm Im}\left(V_{\alpha2}V^\dagger_{l\alpha}
          \lambda'_{jl2}\epsilon_j^*c_\beta\right),\;
 {\rm Im}\left(V_{\alpha2}V^\dagger_{l\alpha}
   \lambda'_{jl2} \epsilon_j^*{M_W\over m_{\chi}}\right)  
< 1.7\times10^{-6} {m_{\chi} m_{\tilde{q}} \over (100 {\rm GeV})^2}\,.
\end{equation}
Collecting all the contributions to the electron and neutron EDMs,
we summarize the resulting  bounds on 
R-parity violating couplings in the Tables.

\section{conclusion}

We have analyzed the electron and neutron EDMs induced by
R-parity violating interactions, taking into
account both the bilinear and trilinear
R-parity violating terms.  It is pointed out that 
the fermion EDMs can  arise at one-loop level as a combined
effect of the bilinear and trilinear terms.
This has to be contrasted 
to the previous analyses showing that nonzero EDMs arise  at two-loop 
level considering only the trilinear R-parity violating vertices.
We have computed the complete one-loop diagrams 
(except those proportional to tiny neutrino masses)
involving both the bilinear and trilinear parameters as presented
in the Appendix.  Among them, the leading contribution comes from the
diagrams with a gauge/top-Yukawa vertex and a R-parity violating trilinear 
vertex for which  the chirality-flip is provided by (bilinear) mass mixing
between R-parity even and odd particles.  These one-loop diagrams
lead to somewhat severe constraints on the products of the R-parity
violating bilinear and trilinear couplings as  summarized in Table I.
If the bilinear R-parity violation is a dominant
source for the neutrino mass $m_{\nu}\sim 0.1$ eV
as was suggested by the Super-Kamionkande
atmospheric neutrino data, the typical upper bounds of 
${\cal O}(10^{-2})$ can be put on some $\lambda_{ijk}$ couplings involving
the first generation index as shown in Table II.

\bigskip

{\bf Acknowledgement}
This work is  supported by Seoam Foundation (K.C.) and 
by grant No.~1999-2-111-002-5 from the interdisciplinary 
Research program of the KOSEF and BK21 project of the
Ministry of Education (K.C. and K.H).



\renewcommand{\arraystretch}{1}
\begin{table}
\begin{tabular}{|c|c|}
\quad\quad\quad\quad\quad\quad\quad\quad coupling 
\quad\quad\quad\quad\quad\quad\quad\quad & upper bound   \\ \hline
$ \lambda_{1j1}\epsilon_j^* c_\beta$    & $1.4\times 10^{-8} x_s$  \\ 
$ \delta^{l,LL}_{13}\lambda_{231}\epsilon_2^* c_\beta$   
                          & $1.4\times 10^{-8} x_s  $  \\ 
$ \delta^{l,LL}_{12}\lambda_{1j2}\epsilon_j^* c_\beta$    
                          & $0.64\times 10^{-7} x_s $  \\ 
$ \delta^{l,LL}_{13}\lambda_{1j3}\epsilon_j^* c_\beta$    
                          & $0.64\times 10^{-7} x_s $  \\ \hline
$ V_{l\alpha}^\dagger V_{\alpha 1} \lambda'_{jl1}\epsilon_j^* c_\beta$   
		 	 & $1.0\times 10^{-7} x_s  $  \\
$ \delta^{l,LL}_{1l}\lambda'_{jl1}\epsilon_j^* c_\beta$   
			 	 & $1.0\times 10^{-7} x_s  $  \\

$ V_{l\alpha}^\dagger V_{\alpha 2} \lambda'_{jl2}\epsilon_j^* c_\beta$   
			 	 & $1.7\times 10^{-6} x_s  $  \\
$ V_{l3}^\dagger \lambda'_{jl1} {B_j^* \over m_{\tilde{l}}^2} $
  & $2.5\times 10^{-6}
	\left({0.01\over V_{31}}{m_{\tilde{l}}\over 100{\rm GeV}}\right) $ 
\end{tabular}

\bigskip

\caption {
Bounds on the imaginart part of the combinations of trilinear and 
bilinear couplings. Note that  $\epsilon_j^* c_\beta$ can be replaced by 
$\epsilon_j^* M_W/m_\chi$. Here  $x_s \equiv {m_\chi m_{\tilde{f}}\over 
(100 {\rm GeV}^2)}$. 
}
\vspace{1cm}
\end{table}

\begin{table}
\begin{tabular}{|c|c|}
\quad\quad\quad\quad\quad\quad\quad\quad coupling 
\quad\quad\quad\quad\quad\quad\quad\quad & upper bound   \\ \hline
$ \lambda_{121},\lambda_{131}$       & $ 0.013 x_s x_\nu$  \\ 
$ \delta^{l,LL}_{13}\lambda_{231}$   & $ 0.013 x_s x_\nu $  \\ 
$ \delta^{l,LL}_{13}\lambda_{123}$   & $ 0.059 x_s x_\nu $  \\ 
$ \delta^{l,LL}_{13}\lambda_{133}$   & $ 0.059 x_s x_\nu $  \\ 
\hline
$ \lambda'_{j11}$                 & $ 0.10 x_s x_\nu $  \\ 
$ \lambda'_{j21}$                 & $ 0.45 x_s x_\nu $  \\ 
$ \delta^{u,LL}_{13}\lambda'_{j31}$  & $ 0.10 x_s x_\nu $  \\ 
\end{tabular}

\bigskip

\caption{
Bounds on the magnitude of single trilinear coupling under the assumption that
neutrino masses come dominantly from the R-parity violating
bilinear terms. As a reference, we take $m_\nu \sim 0.1$ eV and then
$x_\nu \equiv ({F_N\over 100{\rm GeV}})^{-1}
({m_\nu \over 0.1{\rm eV}})^{-\haf}$.\\
}
\end{table}

\appendix

\section{Full one-loop contributions to the fermion dipole moments from
R-parity violation}

In this Appendix, we provide complete expressions for
the one-loop fermion dipole moments induced by R-parity violating couplings.
For this, let us first define some loop functions.
For the basic loop functions given by
\bea
G_f(t) &\equiv& {t-3 \over 2(1-t)^2} - {\ln t \over (1-t)^3} \,,
\nonumber \\
G_s(t) &\equiv& {t+1 \over 2(1-t)^2} + {t \ln t \over (1-t)^3}\,,
\nonumber \eea
the modified loop functions which are propotional to 
the chirality-flipping fermion masses are defined as 
\bea
G_f(m_f;m_\alpha) &\equiv& {m_f \over m_\alpha} 
  G_f\left({m_f^2 \over m^2_\alpha} \right) \,, \nonumber  \\
G_s(m_f;m_\alpha) &\equiv& {m_f \over m_\alpha} 
  G_s\left({m_f^2 \over m^2_\alpha} \right). \nonumber 
\eea
These functions behave well in the limit $m_f\to 0$ and 
mildly depend on $m_f/m_\alpha$ as shown in FIG.~4.
When the mass insertion approximation is used, 
another useful set of (dimensionful) 
loop functions are
\bea
H_f(m_f;m_\alpha, m_\beta) &=& 
{m_f m_\alpha m_\beta \over m_\alpha^2 - m_\beta^2}
\left[ {1\over m_\alpha^2} G_f\left({m_f^2\over m_\alpha^2}\right) 
- {1\over m_\beta^2} G_f\left({m_f^2\over m_\beta^2}\right) \right]
\nonumber \\
H_s(m_f;m_\alpha, m_\beta) &=&
{m_f m_\alpha m_\beta \over m_\alpha^2 - m_\beta^2}
\left[ {1\over m_\alpha^2} G_s\left({m_f^2\over m_\alpha^2}\right) 
- {1\over m_\beta^2} G_s\left({m_f^2\over m_\beta^2}\right) \right]
\nonumber 
\eea
Any contribution to the electric dipole moment between the 
two fermions $f_i$ and $f_k$ can be expressed in the following form:
\bea
d_{f,ik} &=& {e \over 16\pi^2} 
  {1\over 2i}
  \left[ \left( \delta^{f,AB}_{\alpha\beta} \cdot {\cal A}\right)_{ik}
  -\left( \delta^{f,AB}_{\alpha\beta} \cdot {\cal A}\right)_{ki}^*
  \right] 
\eea
where $f$ stands for the external quarks or charged leptons, 
$i,k$ are their generation indices, 
$AB$ denote the chirality of intermediate sfermions, i.e.  $AB=LL,LR,RR$, and 
$\alpha$ and $\beta$ are the indices for
those sfermions.
Here we do not seperately treat the diagrams without 
sfermion flavor mixing.
The expressions for such diagrams 
can be easily obtained from those for the diagrams with
sfermion flavor mixing
by changing $\delta^{f,AB}_{\alpha\beta}, H_f, H_s$ as
follows:
\bea
\delta^{f,AB}_{\alpha\beta} &\rightarrow & \delta_{\alpha\beta} \,,
\nonumber \\
H_f(m_f;m_\alpha, m_\beta) &\rightarrow & 
  {1\over m_\alpha}  G_f(m_f;m_\alpha) \,, \nonumber \\
H_s(m_f;m_\alpha, m_\beta) &\rightarrow &
  {1\over m_\alpha} G_s(m_f;m_\alpha) \nonumber \,.
\eea

The part $(\delta\cdot{\cal A})_{ik}$ for the
charged lepton dipole moments between $l_i$ and $l_k$ 
are then given by
\bea \label{Ae}
\left(\delta\cdot{\cal A} \right)^e_{ik}  
= &+& g\u^R_{W\chi_n} \u^L_{j\chi_n} \delta^{l,LL}_{i\alpha} 
  \lambda_{\alpha j k}
 H_f(m_{\chi^\pm_n}; m_{\tilde{\nu}_i}, m_{\tilde{\nu}_\alpha} ) \\
&-& {1\over \sqrt{2}} \left( g \u^N_{W\chi_n}+ g' \u^N_{B\chi_n} \right) 
 \u^N_{j\chi_n} \delta^{l,LL}_{i\alpha} \lambda_{\alpha j k} 
 H_s(m_{\chi^0_n}; m_{\tilde{e}_i}, m_{\tilde{e}_\alpha} ) \nonumber\\
&+& \sqrt{2} g' \u^N_{B\chi_n} \u^N_{j\chi_n} \delta^{l,RR}_{k\alpha} 
   \lambda_{i j \alpha} 
 H_s(m_{\chi^0_n}; m_{\tilde{e^c}_k}, m_{\tilde{e^c}_\alpha} ) \nonumber\\
&+& g \u^R_{Wk} \delta^{l,LL}_{l\alpha} \lambda_{\alpha i l} 
 H_f(m_{e_l}; m_{\tilde{\nu}_l}, m_{\tilde{\nu}_\alpha} ) \nonumber\\
&-& \haf \left[  h^e_i \left\{
 \left( \u^S_{H_1^R \phi^s_l} +i\u^S_{H_1^I \phi^s_l} \right)
 \left( \u^S_{j^R \phi^s_l} +i\u^S_{j^I \phi^s_l} \right)
{1\over m_{\phi^s_l}} G_f(m_{e_i}; m_{\phi^s_l}) 
 \right\} + (i \leftrightarrow k) \right] \lambda_{jik} \nonumber\\
&+& h^e_i \u^N_{H_1 \chi_n} \u^N_{j \chi_n} \delta^{l,LR}_{\alpha i} 
  \lambda_{j\alpha k}
 H_s(m_{\chi^0_n}; m_{\tilde{e}_\alpha}, m_{\tilde{e^c}_i} ) \nonumber\\
&+& h^e_k \u^N_{H_1 \chi_n} \u^N_{j \chi_n} \delta^{l,LR}_{\alpha k} 
  \lambda_{ji\alpha}
 H_s(m_{\chi^0_n}; m_{\tilde{e^c}_\alpha}, m_{\tilde{e}_i} ) \nonumber\\
&-& g V_{\alpha l}V^{\dagger}_{\gamma\beta} 
 \u^R_{Wk} \delta^{u,LL}_{\alpha\beta} \lambda'_{i\gamma l} \left[ 
 2 H_s(m_{d_l}; m_{\tilde{u}_\alpha}, m_{\tilde{u}_\beta} ) 
 + H_f(m_{d_l}; m_{\tilde{u}_\alpha}, m_{\tilde{u}_\beta} ) \right] 
   \nonumber\\
&+& h^u_\alpha V_{\alpha l}V^{\dagger}_{\gamma\beta} 
 \u^R_{Hk} \delta^{u,RL}_{\alpha\beta} \lambda'_{i\gamma l} \left[ 
 2 H_s(m_{d_l}; m_{\tilde{u^c}_\alpha}, m_{\tilde{u}_\beta} ) 
 + H_f(m_{d_l}; m_{\tilde{u^c}_\alpha}, m_{\tilde{u}_\beta} ) \right] 
   \nonumber\\
&+& h^u_l V_{l\alpha}V^{\dagger}_{\gamma l} 
 \u^R_{Hk} \delta^{d,RL}_{\beta\alpha} \lambda'_{i\gamma \beta} \left[ 
 2 H_s(m_{u_l}; m_{\tilde{d^c}_\gamma}, m_{\tilde{d}_\alpha} ) 
 + H_f(m_{u_l}; m_{\tilde{d^c}_\gamma}, m_{\tilde{d}_\alpha} ) \right]\,,
   \nonumber 
\eea
and for the dipole moments between the down type-quarks $d_i$ and $d_k$,
\bea  \label{Ad}
\left(\delta\cdot{\cal A} \right)_{ik}^d 
= &-&  g \u^L_{j\chi^-_n} \u^R_{W\chi^+_n} 
 V_{\alpha i}V^{\dagger}_{l\beta} \delta^{u,LL}_{\alpha\beta} \lambda'_{jlk} 
  \left[ {2\over 3} H_s(m_{\chi_n}; m_{\tilde{u}_\alpha}, m_{\tilde{u}_\beta} ) 
 + H_f(m_{\chi_n}; m_{\tilde{u}_\alpha}, m_{\tilde{u}_\beta} )  \right] 
        \\
&+& {1\over 3\sqrt{2}} \left( g \u^N_{W\chi_n}- {g'\over 3} \u^N_{B\chi_n} 
  \right) 
 \u^N_{j\chi^0_n} \delta^{d,LL}_{i\alpha} \lambda_{j\alpha k} 
 H_s(m_{\chi^0_n}; m_{\tilde{d}_i}, m_{\tilde{d}_\alpha} ) \nonumber\\
&+& {\sqrt{2}\over 9} g' \u^N_{B\chi_n} \u^N_{j\chi_n} \delta^{d,RR}_{k\alpha} 
 \lambda'_{j i \alpha} 
 H_s(m_{\chi^0_n}; m_{\tilde{d^c}_k}, m_{\tilde{e^c}_\alpha} ) \nonumber\\
&-& h^u_\alpha \u^{C*}_{H_2\phi^-_n} \u^C_{j\phi^-_n} 
 V_{\alpha i}V^{\dagger}_{l\alpha} \lambda'_{jlk} 
 {1\over m_{\phi^-_n}} \left[ 
 {2\over 3} G_f(m_{u_\alpha} ;m_{\phi^-_n}) 
 + G_s(m_{u_\alpha} ;m_{\phi^-_n})  \right] \nonumber\\
&+& h^u_\alpha \u^L_{j\chi^-_n} \u^R_{W\chi^+_n} 
 V_{\alpha i}V^{\dagger}_{l\beta} \delta^{u,LR}_{\beta\alpha} \lambda'_{jlk} 
 \left[ {2\over 3} H_s(m_{\chi_n}; m_{\tilde{u}_\beta}, m_{\tilde{u^c}_\alpha}) 
 +H_f(m_{\chi_n}; m_{\tilde{u}_\beta}, m_{\tilde{u^c}_\alpha} )  \right] 
  \nonumber\\
&-& {1\over 6}\left[ h^d_i \left\{ 
 \left( \u^S_{H_1^R \phi^s_l} +i\u^S_{H_1^I \phi^s_l} \right)
 \left( \u^S_{j^R \phi^s_l} +i\u^S_{j^I \phi^s_l} \right)
{1\over m_{\phi^s_l}} G_f(m_{d_i}; m_{\phi^s_l}) 
 \right\} + (i \leftrightarrow k) \right] \lambda'_{jik} \nonumber\\
&+& {1\over 3}h^d_i \u^N_{H_1 \chi_n} \u^N_{j \chi_n} \delta^{d,LR}_{\alpha j} 
 \lambda_{j\alpha k}
 H_s(m_{\chi^0_n}; m_{\tilde{d}_\alpha}, m_{\tilde{d^c}_i} ) \nonumber\\
&+& {1\over 3}h^d_k \u^N_{H_1 \chi_n} \u^N_{j \chi_n} \delta^{d,RL}_{\alpha k} 
 \lambda_{ji\alpha}
 H_s(m_{\chi^0_n}; m_{\tilde{d^c}_\alpha}, m_{\tilde{d}_k} )\,,\nonumber 
\eea
and finally for the up type quarks $u_i$ and $u_k$,
\bea  \label{Au}
\left(\delta\cdot{\cal A}\right)_{ik}^u =
&-& h^u_k \u^{C*}_{H_2 \phi^-_n} \u^C_{j \phi^-_n} V_{kl} V^\dagger_{\alpha i}
 \lambda'_{i\alpha l} {1\over m_{\phi^-_n}}\left\{ -{1\over 3}
 G_f(m_{d_l}; m_{\phi^-_n}) +G_s(m_{d_l}; m_{\phi^-_n}) \right\} \\
&-& h^u_k \u^L_{j \chi_n} \u^R_{H \chi_n} V_{k\alpha} V^\dagger_{\gamma i}
 \delta^{d,LR}_{\alpha\beta} \lambda'_{j\gamma\beta} \left\{ {1\over 3}
 H_s(m_{\chi^\pm_n}; m_{\tilde{d}_\alpha}, m_{\tilde{d}_\beta} ) 
 - H_f(m_{\chi^\pm_n}; m_{\tilde{d}_\alpha}, m_{\tilde{d}_\beta} ) \right\} 
    \nonumber 
\eea
Here $\phi^-_n$, $\phi^s_l$ denotes the charged and neutral scalar 
mass-eigenstates, respectively.

\begin{figure}
\begin{center}
\begin{picture}(300,170)(0,0)

\ArrowLine(10,50)(70,50)
\ArrowLine(150,50)(70,50)
\Line(145,45)(155,55) 
\Line(155,45)(145,55)
\ArrowLine(150,50)(230,50)
\ArrowLine(290,50)(230,50)

\DashArrowArcn(150,50)(80,90,0){5}
\DashArrowArcn(150,50)(80,180,90){5}
\Vertex(150,130){4}

\Text(40,38)[]{$e_i$}
\Text(110,38)[]{$\tilde{W}^+$}
\Text(190,38)[]{$e_j$}
\Text(260,38)[]{$e_k^c$}

\Text(75,115)[]{$\tilde{\nu}_i$}
\Text(225,115)[]{$\tilde{\nu}_\alpha$}

\Text(70,38)[]{$g$}
\Text(230,38)[]{$\lambda_{\alpha jk}$}

\Text(150,10)[]{\rm (a)}
\end{picture}
\end{center}

\begin{center}
\begin{picture}(300,170)(0,0)

\ArrowLine(10,50)(70,50)
\ArrowLine(150,50)(70,50)
\Line(145,45)(155,55) 
\Line(155,45)(145,55)
\ArrowLine(150,50)(230,50)
\ArrowLine(290,50)(230,50)

\DashArrowArcn(150,50)(80,90,0){5}
\DashArrowArcn(150,50)(80,180,90){5}
\Vertex(150,130){4}

\Text(40,38)[]{$e_i$}
\Text(110,38)[]{$\tilde{B},\tilde{W}_3$}
\Text(190,38)[]{$\nu_j$}
\Text(260,38)[]{$e_k^c$}

\Text(75,115)[]{$\tilde{e}_i$}
\Text(225,115)[]{$\tilde{e}_\alpha$}

\Text(70,38)[]{$g',g$}
\Text(230,38)[]{$-\lambda_{\alpha jk}$}

\Text(150,10)[]{\rm (b)}
\end{picture}
\end{center}

\begin{center}
\begin{picture}(300,170)(0,0)

\ArrowLine(10,50)(70,50)
\ArrowLine(150,50)(70,50)
\Line(145,45)(155,55) 
\Line(155,45)(145,55)
\ArrowLine(150,50)(230,50)
\ArrowLine(290,50)(230,50)

\DashArrowArc(150,50)(80,0,90){5}
\DashArrowArc(150,50)(80,90,180){5}
\Vertex(150,130){4}

\Text(40,38)[]{$e_i$}
\Text(110,38)[]{$\nu_j$}
\Text(190,38)[]{$\tilde{B}$}
\Text(260,38)[]{$e_k^c$}

\Text(75,115)[]{$\tilde{e}^c_\alpha$}
\Text(225,115)[]{$\tilde{e}^c_k$}

\Text(70,38)[]{$-\lambda_{ij\alpha}$}
\Text(230,38)[]{$g'$}

\Text(150,10)[]{\rm (c)}
\end{picture}
\end{center}
\caption{Most important diagrams for the charged lepton EDM.  
The black blobs denote possible scalar mixing and 
the crosses are chirality-fliping mass insertions.}
\end{figure}

\begin{figure}
\begin{center}
\begin{picture}(300,170)(0,0)

\ArrowLine(10,50)(70,50)
\ArrowLine(150,50)(70,50)
\Line(145,45)(155,55) 
\Line(155,45)(145,55)
\ArrowLine(150,50)(230,50)
\ArrowLine(290,50)(230,50)

\DashArrowArcn(150,50)(80,90,0){5}
\DashArrowArcn(150,50)(80,180,90){5}
\Vertex(150,130){4}

\Text(35,38)[]{$d_i$}
\Text(110,38)[]{$\tilde{W}^+$}
\Text(190,38)[]{$e_j$}
\Text(270,38)[]{$d_k^c$}

\Text(75,115)[]{$\tilde{u}_\alpha$}
\Text(225,115)[]{$\tilde{u}_\beta$}

\Text(70,38)[]{$g V_{\alpha i}$}
\Text(230,38)[]{$\lambda'_{jlk}V^\dagger_{l\beta}$}

\Text(150,10)[]{\rm (a)}
\end{picture}
\end{center}

\begin{center}
\begin{picture}(300,170)(0,0)

\ArrowLine(10,50)(70,50)
\ArrowLine(150,50)(70,50)
\Line(145,45)(155,55) 
\Line(155,45)(145,55)
\ArrowLine(150,50)(230,50)
\ArrowLine(290,50)(230,50)

\DashArrowArc(150,50)(80,0,90){5}
\DashArrowArc(150,50)(80,90,180){5}
\Vertex(150,130){4}

\Text(40,38)[]{$d_i$}
\Text(110,38)[]{$\nu_j$}
\Text(190,38)[]{$\tilde{B}$}
\Text(260,38)[]{$d_k^c$}

\Text(75,115)[]{$\tilde{d}^c_\alpha$}
\Text(225,115)[]{$\tilde{d}^c_k$}

\Text(70,38)[]{$\lambda'_{ji\alpha}$}
\Text(230,38)[]{$g'$}

\Text(150,10)[]{\rm (b)}
\end{picture}
\end{center}

\begin{center}
\begin{picture}(300,170)(0,0)

\ArrowLine(10,50)(70,50)
\ArrowLine(150,50)(70,50)
\Line(145,45)(155,55) 
\Line(155,45)(145,55)
\ArrowLine(150,50)(230,50)
\ArrowLine(290,50)(230,50)

\DashArrowArcn(150,50)(80,90,0){5}
\DashArrowArc(150,50)(80,90,180){5}
\Line(145,135)(155,125)
\Line(145,125)(155,135)

\Text(35,38)[]{$d_i$}
\Text(110,38)[]{$t^c$}
\Text(190,38)[]{$t$}
\Text(270,38)[]{$d_k^c$}

\Text(75,115)[]{$H_2^+$}
\Text(225,115)[]{$\tilde{e}_j$}

\Text(70,38)[]{$h_t V_{3i}$}
\Text(230,38)[]{$\lambda'_{jlk}V^\dagger_{l3}$}

\Text(150,10)[]{\rm (c)}
\end{picture}
\end{center}
\caption{Most important diagrams for the down-type quark EDM.
         There exist also diagrams similar to FIG.~1b,
	 and also a supersymmetric counterpart of (c).}
\end{figure}

\begin{figure}

\begin{center}
%
%
%
%
%
%
%
\begin{picture}(300,170)(0,0)

\ArrowLine(10,50)(70,50)
\ArrowLine(150,50)(70,50)
\Line(145,45)(155,55) 
\Line(155,45)(145,55)
\ArrowLine(150,50)(230,50)
\ArrowLine(290,50)(230,50)

\DashArrowArcn(150,50)(80,90,0){5}
\DashArrowArc(150,50)(80,90,180){5}
\Line(145,135)(155,125)
\Line(145,125)(155,135)

\Text(40,38)[]{$u_i$}
\Text(110,38)[]{$e_j$}
\Text(190,38)[]{$\tilde{H}^+_2$}
\Text(260,38)[]{$u_k^c$}

\Text(75,115)[]{$\tilde{d}^c_\alpha$}
\Text(225,115)[]{$\tilde{d}_k$}

\Text(70,38)[]{$\lambda'_{ji\alpha}$}
\Text(230,38)[]{$h^u_k$}

\end{picture}
\end{center}
\caption{Leading diagram for the up-type quark EDM. There is also
       a  supersymmetric counterpart of this.}


\begin{center}
\begin{picture}(360,300)(0,0)
\SetOffset(30,60)
\LinAxis(0,0)(300,0)(3,10,5,0,1.3)
\LinAxis(0,0)(0,200)(1,10,5,0,1.3)
\Line(300,0)(300,200)
\Line(0,200)(300,200)
\Text(0,-10)[]{0}
\Text(100,-10)[]{1}
\Text(200,-10)[]{2}
\Text(270,-25)[]{$t = {m_f\over m_\phi}$}
\Text(-20,200)[]{1.0}
\Text(-20,160)[]{0.8}
\Text(-20,120)[]{0.6}
\Text(-20,080)[]{0.4}
\Text(-20,040)[]{0.2}
\Text(280,44)[]{$G_f$}
\Text(280,13)[]{$G_s$}
\SetScale{100.}
\SetWidth{0.005}
\Curve{( .0001, .0001)
( .0021, .0021)
( .0041, .0041)
( .0061, .0061)
( .0081, .0081)
( .0101, .0101)
( .0601, .0583)
( .1101, .1020)
( .1601, .1404)
( .2101, .1738)
( .2601, .2025)
( .3101, .2270)
( .3601, .2477)
( .4101, .2651)
( .4601, .2797)
( .5101, .2919)
( .5601, .3019)
( .6101, .3101)
( .6601, .3167)
( .7101, .3219)
( .7601, .3259)
( .8101, .3289)
( .8601, .3311)
( .9101, .3324)
( .9601, .3332)
(1.0100, .3333)
(1.2100, .3297)
(1.4100, .3218)
(1.6100, .3117)
(1.8100, .3005)
(2.0100, .2890)
(2.2100, .2777)
(2.4100, .2666)
(2.6100, .2561)
(2.8100, .2460)}
\Curve{( .0001, .0034)
( .0021, .0455)
( .0041, .0779)
( .0061, .1061)
( .0081, .1318)
( .0101, .1554)
( .0401, .3978)
( .0701, .5443)
( .1001, .6444)
( .1301, .7155)
( .1601, .7666)
( .1901, .8033)
( .2201, .8292)
( .2501, .8469)
( .2801, .8583)
( .3101, .8647)
( .3401, .8672)
( .3701, .8666)
( .4001, .8635)
( .4301, .8586)
( .4601, .8521)
( .4901, .8443)
( .5001, .8416)
( .6001, .8095)
( .7001, .7736)
( .8001, .7368)
( .9001, .7009)
(1.0100, .6633)
(1.2100, .6012)
(1.4100, .5474)
(1.6100, .5009)
(1.8100, .4608)
(2.0100, .4260)
(2.2100, .3956)
(2.4100, .3690)
(2.6100, .3455)
(2.8100, .3247)
}
\SetScale{1.}
\SetWidth{0.5}
\end{picture}
\end{center}

\caption{Loop functions $G_f(m_f;m_\alpha)$ and $G_s(m_f;m_\alpha)$ 
as a function of $t = m_f/m_\alpha$}

\end{figure}


\begin{thebibliography}{99}
%
\def\plb#1#2#3{Phys.\ Lett.\       {\bf B#1}, #2 (#3)}
\def\npb#1#2#3{Nucl.\ Phys.\       {\bf B#1}, #2 (#3)}
\def\prd#1#2#3{Phys.\ Rev.\        {\bf D#1}, #2 (#3)}
\def\prl#1#2#3{Phys.\ Rev.\ Lett.\ {\bf #1},  #2 (#3)}
\def\mpl#1#2#3{Mod.\ Phys.\ Lett.\ {\bf A#1}, #2 (#3)}
\def\rep#1#2#3{Phys.\ Rep.\        {\bf #1},  #2 (#3)}
\def\sci#1#2#3{Science             {\bf #1},  #2 (#3)}
\def\astro#1#2#3{Astrophys.\ J.\   {\bf #1},  #2 (#3)}
%
\bibitem{eedm}
K.~Abdullah {\it et al.}, \prl {65}{2340}{1990};
E.~Commins {\it et al.}, Phys.~Rev.~{\bf A~50} (1994) 2960;
B.E.~Sauer, J.~Wang and E.A.~Hinds, \prl{74}{1554}{1995};
{\it J. Chem. Phys} {\bf 105} (1996) 7412.

\bibitem{nedm}
P.G.~Harris {\it et al.}, \prl{82}{904}{1999}.

%

\bibitem{mssm} 
J. Ellis, S. Ferrara and D.V. Nanopoulos, \plb{114}{231}{1982};
W. Buchmuller and D. Wyler, \plb{121}{321}{1983};
J. Polchinski and M.B. Wise, \plb{125}{393}{1983};
J-M.~Gerard, {\it et al.}, \npb{253}{93}{1985}; 
M.~Dugan, B.~Greenstein and L.~Hall, \npb{255}{413}{1985};
A. Sanda, \prd{32}{2992}{1985}.

\bibitem{cancellation} 
T.~Ibrahim and P.~Nath, \plb{418}{98}{1998}; \prd{57}{478}{1998};
T.~Falk and K.A.~Olive, \plb{439}{71}{1998}; 
M.~Brhlik, G.J.~Good and G.L.~Kane, \prd{59}{115004}{1999}; 
S.~Pokorski, J.~Rosiek and C.A.~Savoy, \npb{570}{81}{2000}.

\bibitem{symmetry} 
V. Ben-Hamo and Y. Nir, \plb{339}{77}{1994};
T. Banks, {\it et al.}, \prd{52}{5319}{1995};
E. J. Chun and A. Lukas, \plb{387}{99}{1996};
K. Choi, E. J. Chun and H. Kim, \prd{55}{7010}{1997}; 
K. Choi, E. J. Chun and H. Kim, \plb{394}{89}{1997};
G. Eyal and Y. Nir, JHEP 9906, 025 (1999);
Q. Shafi and Z. Tavartkiladze, hep-ph/9909238.

\bibitem{rpnumass} 
L.  Hall and M. Suzuki, \npb{231}{419}{1984};
F. M. Borzumati, {\it et al.}, Phys. Lett.  {\bf B384} (1996) 123; 
B. de Carlos and P. L. White, Phys. Rev.  {\bf D54} (1996) 3427; 
A. Yu. Smirnov and  F. Vissani, Nucl. Phys. {\bf B460} (1996) 37; 
R. Hempfling, Nucl. Phys {\bf B478} (1996) 3; 
H. P. Nilles and N. Polonsky, Nucl. Phys. {\bf B484} (1997) 33; 
E. Nardi, Phys. Rev. {\bf D55} (1997) 5772;
E. J. Chun, {\it et al.}, \npb{544}{89}{1999};
A. S. Joshipura and S. K. Vempati, \prd{60}{095009}{1999};
J. Ferrandis, \prd{60}{095012}{1999};
M. Bisset, {\it et al.}, hep-ph/9811498;
S. Rakshit, G. Bhattacharyya, \prd{59}{091701}{1999};
R. Adhikari, G. Omanovic, \prd{59}{073003}{1999};
K. Choi, E. J. Chun and K. Hwang, \prd{60}{031301}{1999};
D. E. Kaplan and A. Nelson,  hep-ph/9901254;
A. Abada and M. Losada, hep-ph/9908352;
J. L. Chkareuli, {\it et al.}, hep-ph/9908451;
J. C. Romao, {\it et al.}, hep-ph/9907499;
O. Haug, {\it et al.}, hep-ph/9909318.


\bibitem{tata}
R.M.~Godbole, S.~Pakvasa, S.D.~Rindani and X.~Tata, hep-ph/9912315

\bibitem{dreiner} 
S.A.~Abel, A.~Dedes and H.K.~Dreiner, hep-ph/9912429

\bibitem{barr} 
S.M. Barr, E.M. Friere and A. Zee, \prl{65}{2626}{1990}.



\bibitem{ptri}
I. Hinchliffe and T. Kaeding, \prd{47}{279}{1993};
A.Y. Simrnov and F. Vissani, \plb{380}{317}{1996}.

\bibitem{pbi}
B. Mukhopadhyaya, {\it et al.}, \plb{443}{191}{1998};
E.J. Chun and J.S. Lee, \prd{60}{075006}{1999}.


\bibitem{kang}
E.J. Chun and S.K. Kang, \prd{61}{078012}{2000},
[hep-ph/9909429].

\bibitem{yuval} 
Y. Grossman and H.E. Haber, \prd{59}{093008}{1999}.

\bibitem{haber}
For a review, see H. Haber and G. Kane, \rep{117}{75}{1985}.

\bibitem{nowa}
A.S. Joshipura and M. Nowakowski, \prd{51}{2421}{1995};
M. Nowakowski and A. Pilaftsis, \npb{461}{19}{1996}.

\bibitem{skatm}  
    The Super-Kamiokande Collaboration, Y. Fukuda, {\it et al.},
    \prl{81}{1562}{1998}.

\bibitem{solan}
For recent analyses, see, 
J.N. Bahcall, P.I. Krastev and A.Y. Smirnov, \prd{58}{096016}{1998}.

\bibitem{scalmix}
F. Gabbiani, {\it et al.}, \npb{477}{321}{1996};
M. Misiak, S. Pokorski and J. Rosiek, in ``Heavy Flavours II'',
{\it ed} A.J. Buras and M. Lindner (World Scientific, 1998),
[hep-ph/9703442].

\bibitem{rgedm}
E. Braaten, C.S. Li and T.C. Yuan, \prl{64}{1709}{1990};
R. Arnowitt, J. Lopez and D.V. Nanopoulos, \prd{42}{2423}{1990}.

\bibitem{cdm}
 X.G. He, B. McKeller and S. Pakvasa,
  Int.\ J. Mod.\ Phys.\ {\bf A4}, 5011 (1989); 
  Erratum, ibid.\ {\bf A6},1063 (1991)

\bibitem{pak}
 X.G. He, B. McKeller and S. Pakvasa, \plb{254}{231}{1991}.


\end{thebibliography}
\end{document}